\documentclass[aps, prl, amsmath, amssymb,preprint, floatfix, lengthcheck, superscriptaddress ]{revtex4-1}
\pdfoutput=1

\usepackage{graphicx}
\usepackage[margin=0.95in]{geometry}
\usepackage{amsmath}
\usepackage{hyperref}
\usepackage{multirow}

\newcommand{\sub}[1]{\ensuremath{_{\textrm{#1}}}} \newcommand{\super}[1]{\ensuremath{^{\textrm{#1}}}} \newcommand{\rmd}{\ensuremath{\mathrm{d}}}

\newcommand{\HarvardSEAS}{John A. Paulson School of Engineering and Applied Sciences, Harvard University, Cambridge, MA, USA}

\newcommand{\RPIMSE}{Department of Materials Science and Engineering, Rensselaer Polytechnic Institute, Troy, NY, USA}

\usepackage[usenames,dvipsnames]{xcolor}

\begin{document}

\title{Microscopic Origins of Hydrodynamic Transport in Type-II Weyl Semimetal WP\sub{2}}

\bibliographystyle{apsrev4-1}
\author{Jennifer Coulter} \email{jcoulter@g.harvard.edu}\affiliation{\HarvardSEAS}
\author{Ravishankar Sundararaman} \affiliation{\RPIMSE}
\author{Prineha Narang} \email{prineha@seas.harvard.edu}\affiliation{\HarvardSEAS}
\date{\today}

\begin{abstract}
The origins of hydrodynamic transport in strongly interacting Dirac and Weyl semimetals have remained elusive in theoretical descriptions and experimental measurements. We investigate the structure and microscopic properties of transport in WP\sub{2}, a type-II Weyl semimetal, to probe the emergence of hydrodynamic phenomena. We characterize the quantum behavior underlying the hydrodynamic transport regime as a function of temperature through \emph{ab initio} calculations of the relevant microscopic scattering processes, including electron-phonon, electron-electron, and phonon-mediated electron-electron lifetimes. We present a fundamentally new approach to calculate phonon-drag, a mechanism that is invoked in numerous recent experiments, and remains the subject of active debate in the field. Further, we show unique and unexpected features of the lifetime-resolved Fermi surfaces of WP\sub{2} in the hydrodynamic regime and quantify the degree of anisotropy in electron and hole pockets. This description of the microscopic dynamics in hydrodynamic systems like WP\sub{2} indicates the importance of electron-phonon interactions in understanding connections between transport in hydrodynamic materials and strongly correlated quantum systems including unconventional metals and high $T_c$ superconductors.
\end{abstract}

\maketitle

Charge transport is ubiquitous in several fields of physics and underlies key phenomena in condensed matter physics across length scales. Conventionally, transport in materials can be described by tracking the motion of quasiparticles. However, new classes of quantum materials have exhibited unique and mysterious signatures not described by well-established models of scattering and which result in strong violation of the Wiedemann-Franz law. \cite{Levitov:2016fk, gooth_electrical_2017, principi_violation_2015, mahajan_non-fermi_2013} In systems with strong interactions, such as graphene near charge neutrality, the Fermi liquid paradigm of freely propagating coherent quasiparticles can break down. 

Recent experiments have demonstrated that strong interactions between quasiparticles in Dirac and Weyl semimetals can lead to electronic transport described as a viscous hydrodynamic flow in 2D. \cite{PhysRevLett.106.156601, PhysRev.109.1486,  mahajan_non-fermi_2013} 
The collective behavior in these systems results from the rapid exchange of momenta and energy in two-body collisions. Since momentum and energy remain conserved quantities collectively, they give rise to hydrodynamic transport modes. In the language of scattering, this implies momentum-conserving (MC) scattering occurs far more often than momentum-relaxing (MR) processes, resulting in a separation of MC and MR timescales. Since MR scattering occurs very slowly relative to MC scattering, momentum is considered to be `quasi-conserved,' and the relevant conservation equations of hydrodynamics can be used to describe transport. 

While these effects have been studied extensively in graphene \cite{lucas_hydrodynamics_2018, Levitov:2016fk,kumar_superballistic_2017,bandurin2016negative}, recent experimental studies have revealed similar signatures in layered metal PdCoO\sub{2} \cite{moll_evidence_2016}, and more recently in Dirac and Weyl semimetals PtSn\sub{4}\cite{fu_thermoelectric_2018} and the type-II Weyl semimetal tungsten phosphide (WP\sub{2}). In contrast to type-I Weyl semimetals, type-II materials feature `tilted' Weyl cones\cite{soluyanov_type-ii_2015}, which result in a non-vanishing density of states at $E_F$. These materials exhibit a variety of exotic quantum phenomena \cite{Hirschberger:2016rt, moller_type-ii_2017}, including superconductivity, large anisotropic magnetoresistance, and recently, in WP\sub{2}, viscous electron transport. The interplay between topological properties and electronic interactions has been the focus of recent work on this material\cite{Garate_2015}. However, the link between these properties and hydrodynamic transport remains poorly understood.

Though a number of experimental techniques exist to study hydrodynamic transport properties\cite{gooth_electrical_2017,moll_evidence_2016,Levitov:2016fk}, questions still remain in regard to the microscopic interactions that drive a transition to the hydrodynamic regime. Ongoing discussion centers around the role of crystal anisotropy in materials like WP\sub{2}, PdCoO\sub{2} and PtSn\sub{4},\cite{gooth_electrical_2017,moll_evidence_2016,fu_thermoelectric_2018} the impact of phonon-drag and other thermoelectric effects, and the nature of electron-phonon interactions, including the potential importance of of small-angle electron-phonon scattering\cite{fu_thermoelectric_2018}. To address these critical questions, in this \emph{Letter} we present a direct evaluation of the microscopic underpinnings of hydrodynamic flow in WP\sub{2} using \emph{ab initio} calculations of the electron and phonon states to study electron-phonon coupling in the material. From these calculations we quantitatively evaluate relevant scattering rates in WP\sub{2} including those for electron-electron and electron-phonon processes in addition to the momentum relaxing scattering rates measured in experiment. By using WP\sub{2} as a model case, we present a fundamentally new understanding of the critical role of electron-phonon scattering in hydrodynamic transport.  

WP\sub{2} forms a bulk orthorhombic (Cmc2\sub{1}) structure and manifests four pairs of type-II Weyl points in the $k_z = 0$ plane.\cite{autes_robust_2016,razzoli_stable_2018}
In experimental transport measurements, WP\sub{2} is considered quasi-2D as a result of strong magnetotransport anisotropy which favors flow in the a-c crystal plane,
similar to anisotropy observed in PdCoO\sub{2} and PtSn\sub{4}.\cite{gooth_electrical_2017} 

\begin{figure}
\includegraphics[width=\columnwidth]{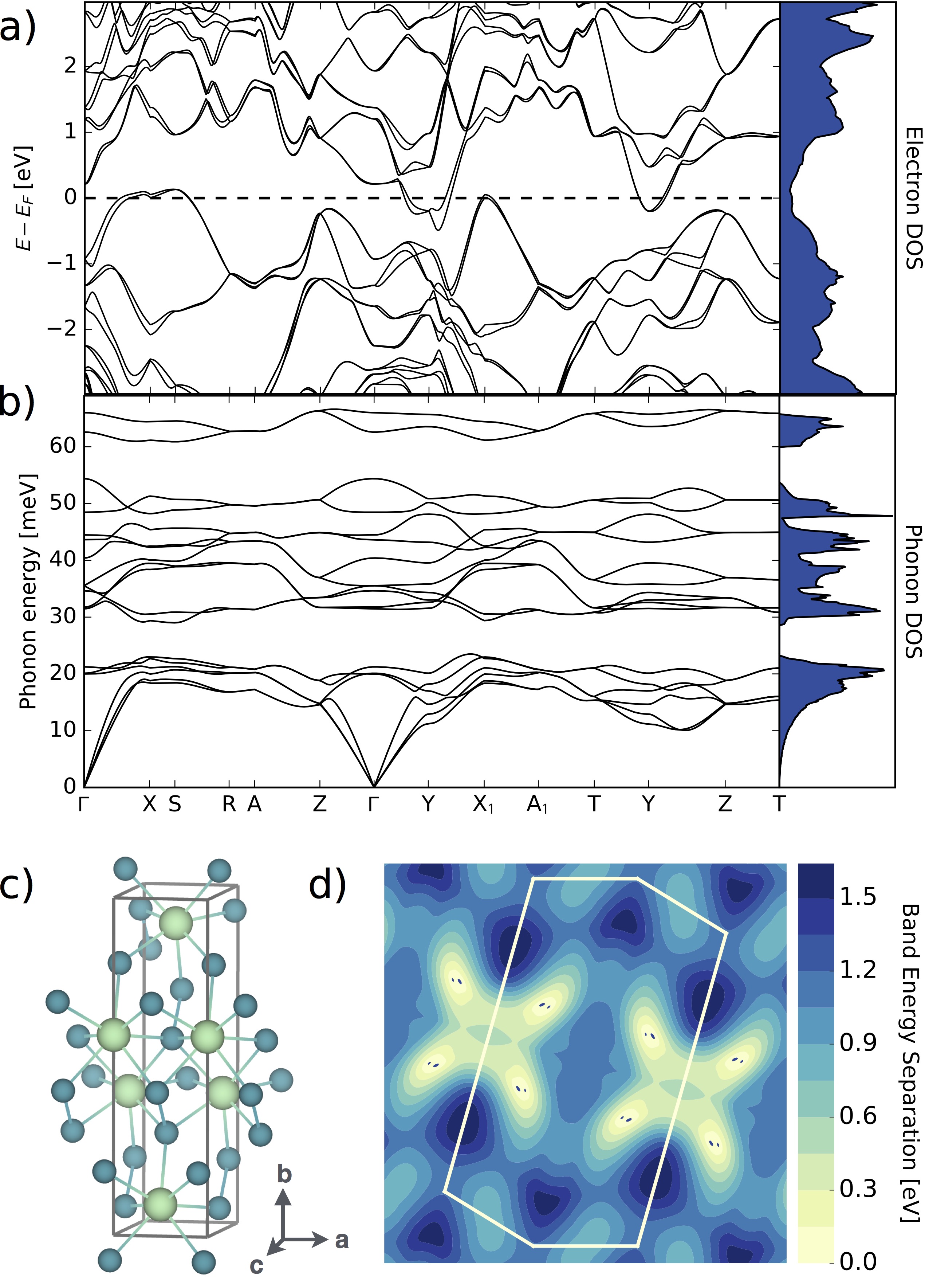}
\caption{\textbf{Structure, dispersions and topology of type-II Weyl semimetal WP\sub{2}.}
(a) Electronic band structure and (b) phonon dispersion, with corresponding densities of states,
for the orthorhombic WP\sub{2} structure shown in (c) calculated using fully-relativistic
\emph{ab initio} calculations (including self-consistent spin-orbit coupling).
(d) Four pairs of Weyl points emerge from electronic structure predictions.
Here, the contour levels indicate the energy separation of the two crossing bands,
with areas of separation less than 0.02 eV in dark blue to identify the Weyl points.
\label{fig:1}}
\end{figure}

Fig.~\ref{fig:1} shows the calculated electron and phonon band structures,
crystal structure and indicates the location of Weyl points in the Brillouin zone.
In our calculations, we locate a pair of Weyl points at $[k_x,k_y] = [-0.282,0.302]$ and $[-0.238,0.283]$ $~1/\AA$ at energies of -0.504 and -0.301 eV respectively, in agreement with previous reports,\cite{razzoli_stable_2018,autes_robust_2016}
with all eight Weyl points accounting for four in-plane symmetries visible
in the contour map of Fig. \ref{fig:1}(d).

From these electron and phonon states, we calculate electron-phonon and electron-electron scattering rates 
($2\hbar\times$ corresponding imaginary part of self-energy) using
Fermi's golden rule and the random-phase approximation respectively,
following our previously established methodology.\cite{NatCom,PhononAssistedDecay,TAparameters,TAanalysis}
We calculate net transport coefficients including resistivities and mobilities based on the
linearized Boltzmann equation with an all-bands relaxation-time approximation.\cite{PhononAssistedDecay}
Additionally, in this work, we estimate the rate of electron-electron scattering
mediated by phonons, rather than the conventional Coulomb-mediated scattering.
Within the random-phase approximation, we find
\begin{multline}
\left(\tau\sub{ee}\super{ph}\right)^{-1}
= \frac{\pi\beta\hbar^2}{2g(\varepsilon_F)}
	\sum_{\alpha} \int \frac{\Omega\rmd\vec{q}}{(2 \pi)^3} G_{\vec{q}\alpha}^2 \\
		\times\int_{-\infty}^{\infty}
			\frac{\omega^2\rmd\omega}{\left|\bar{\omega}_{\vec{q} \alpha} - \omega\right|^2 \sinh^2\frac{\beta\hbar\omega}{2}},
\end{multline}
where $\bar{\omega}_{\vec{q} \alpha} = \omega_{\vec{q} \alpha}(1 + i\pi G_{\vec{q}\alpha})$
are complex phonon frequencies including the linewidth due to phonon-electron scattering,
$\beta = (k_BT)^{-1}$, $g(\varepsilon_F)$ is the density of states at the Fermi level.
Above,
\begin{multline}
G_{\vec{q}\alpha} \equiv \sum_{ab} \int \frac{\Omega \rmd \vec{k}}{(2 \pi)^3}
	\left|g_{\vec{k}a,(\vec{k}+\vec{q}) b}^{\vec{q} \alpha}\right|^2\\
	\times \delta (\varepsilon_{\vec{k}a} -\varepsilon_F)
	\delta (\varepsilon_{(\vec{k}+\vec{q})b} - \varepsilon_F),
\end{multline}
is the weight of each phonon mode in the Eliashberg spectral function,
calculated from first-principles electron energies $\varepsilon_{\vec{k}a}$
and electron-phonon matrix elements $g_{\vec{k}a,\vec{k}'b}^{\vec{q} \alpha}$
(see Supplemental Information for a detailed derivation).
To capture the strong spin-orbit effects in WP\sub{2}, all quantities are derived from relativistic
\emph{ab initio} calculations with self-consistent spin-orbit coupling and
fully account for detailed band structures over the entire Brillouin zone,
implicitly treating processes such as Umklapp scattering.\cite{PhononAssistedDecay}

\begin{figure}[h]
\includegraphics[width=0.8\columnwidth]{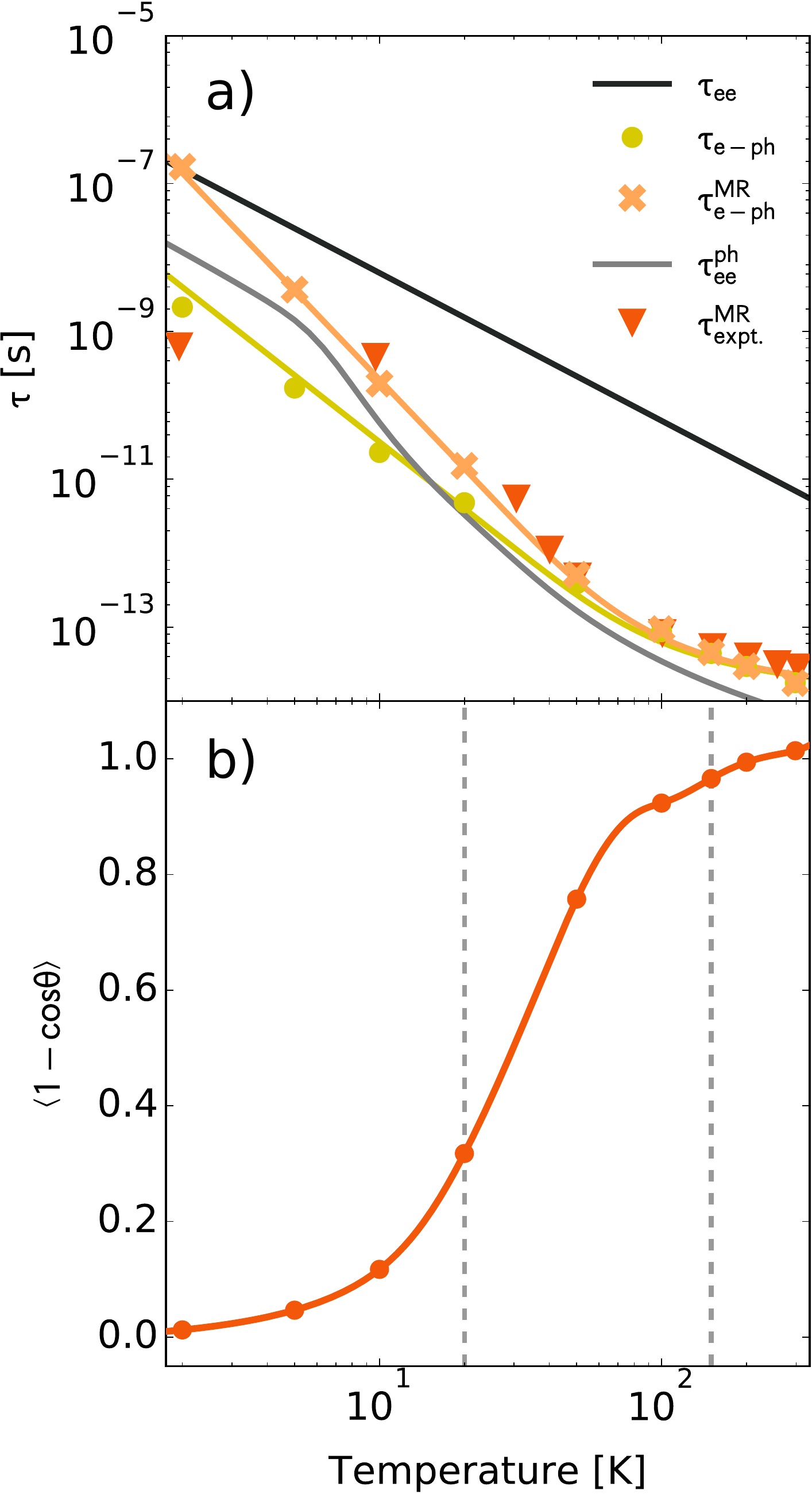}
\caption{\textbf{Predicted temperature dependent lifetimes to distinguish the momentum-relaxing and momentum conserving processes.} (a) \emph{Ab initio} calculations reveal that momentum-relaxing
electron-phonon (e-ph) scattering time $\tau\sub{e-ph}\super{MR}$,
in excellent agreement with experimental measurements $\tau\super{MR}\sub{expt}$,\cite{gooth_electrical_2017}
is much longer than net e-ph scattering time $\tau\sub{e-ph}$.
Coulomb-mediated electron-electron (ee) scattering time $\tau\sub{ee}$
is much longer than the phonon-mediated ee scattering time $\tau\sub{ee}\super{ph}$.
Phonons determine both the momentum-conserving and momentum-relaxing
processes in WP\sub{2}, in stark contrast to previous hydrodynamic
materials like graphene.
(b) Probability of momentum-relaxation by e-ph scattering
$\langle 1-\cos\theta\rangle$ drops off below 100~K
(small-angle scattering increases), providing a pathway
to electronic hydrodynamic behavior. Dashed lines denote where experiment ~\cite{gooth_electrical_2017} reports the onset of a ``hybrid'' ballistic/hydrodynamic regime at 150K and ``pure'' hydrodynamics at 20K. 
\label{fig:2}}
\end{figure}

Hydrodynamic transport requires MC scattering rates to dominate over MR scattering rates
(i.e. correspondingly $\tau\super{MC} \ll \tau\super{MR}$),
conventionally ascribed to electron-electron (ee) and
electron-phonon (e-ph) scattering respectively.
Using our \emph{ab initio} calculations, we can directly evaluate the
contributions of various microscopic mechanisms to MC and MR scattering,
as shown in Fig.~\ref{fig:2}(a).
Note that the calculated electron-phonon MR scattering time $\tau\super{MR}\sub{e-ph}$
is in excellent agreement with experimental estimates.\cite{gooth_electrical_2017}
However, contrary to prior understanding of hydrodynamic transport, the predicted Coulomb electron-electron scattering time is much longer over most of the temperature range, indicating that it cannot be the mechanism
of MC scattering responsible for hydrodynamic behavior in WP\sub{2}.

\begin{figure}
\includegraphics[width=0.8\columnwidth]{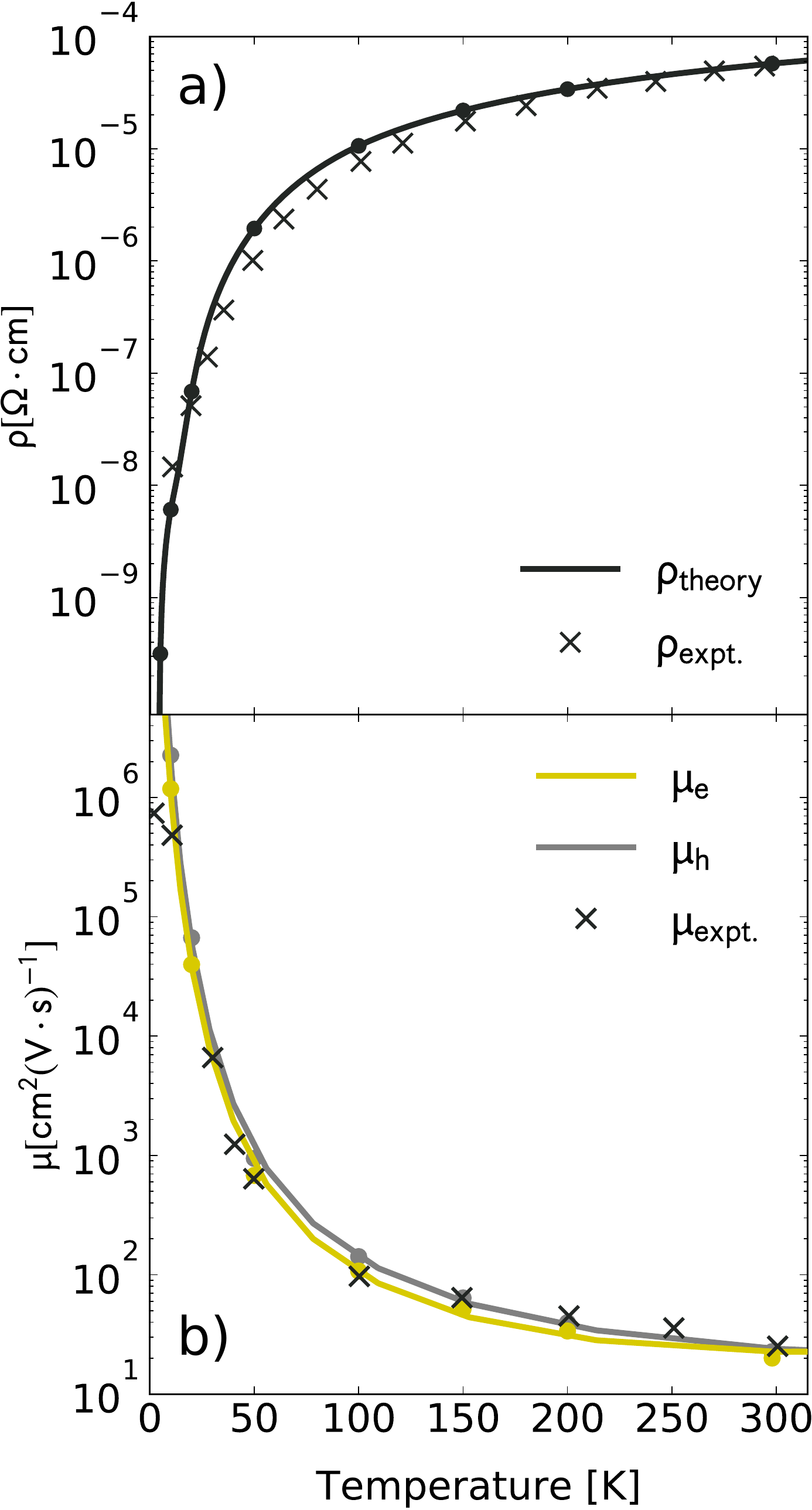}
\caption{\textbf{Transport from first-principles.}
\emph{Ab initio} calculation of the conductivity tensor from a linearized Boltzmann equation
with a full-band relaxation-time approximation predicts (a) resistivity and (b) mobilities,
in excellent agreement with experiment.\cite{schonemann_fermi_2017,gooth_electrical_2017}
This agreement establishes the reliability of first-principles predictions of complex
and highly anisotropic electron-phonon interactions for describing transport properties
in this Weyl semi-metal.
\label{fig:3}}
\end{figure}

\begin{figure*}
\includegraphics[width=0.85\textwidth]{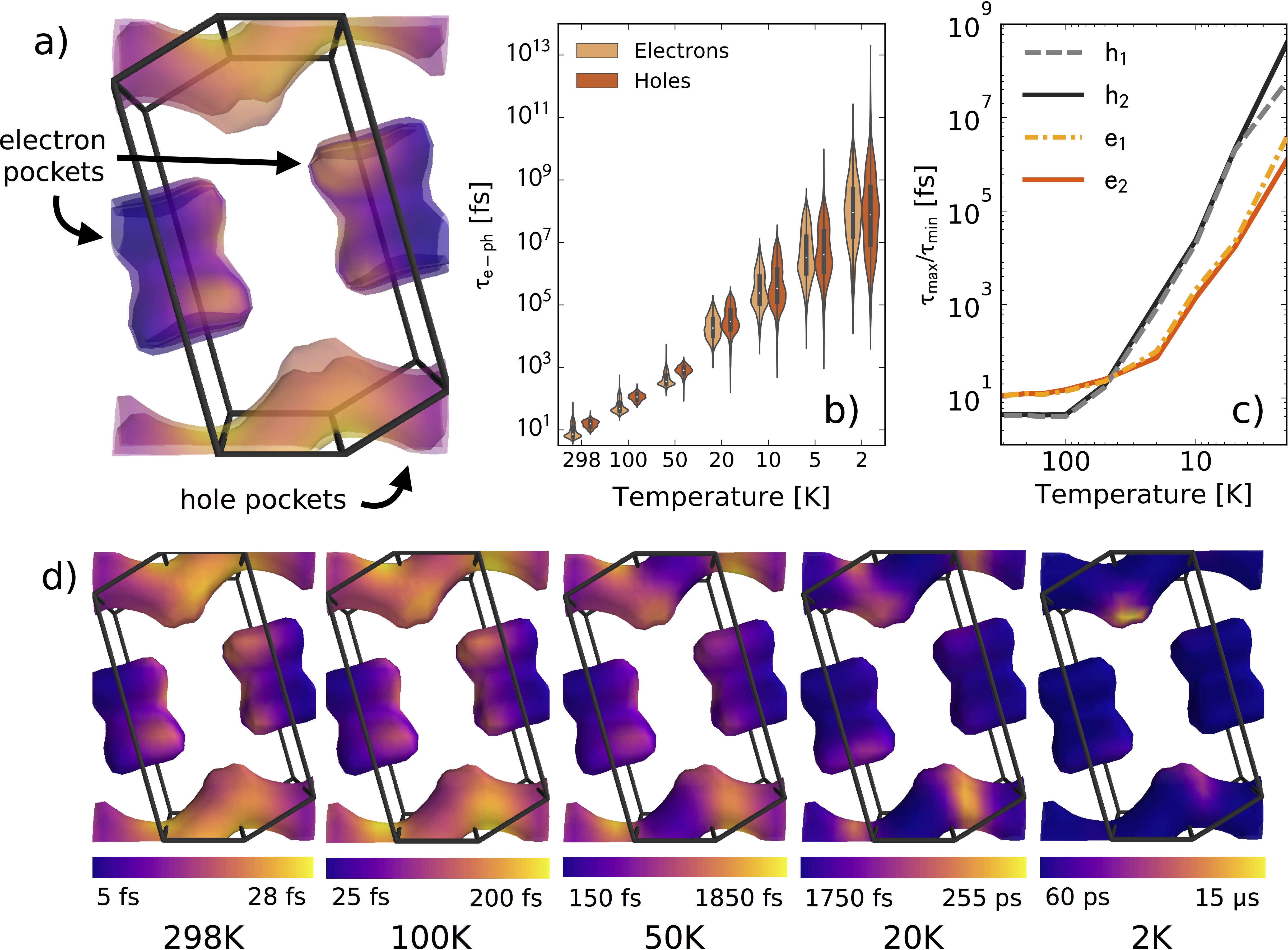}
\caption{\textbf{Electron-phonon interactions at E\sub{F} with scattering rates that are strongly anisotropic on the Fermi surfaces.} The degree of anisotropy, and especially the difference between bowtie-shaped electron pockets and tube-shaped hole pockets strongly increases with decreasing temperature, as shown here. In a) we show that there are two bands which appear one inside the other as a result of spin-orbit coupling, corresponding to each of the hole and electron pockets; in b), we show the distribution of lifetime values for the bands associated with the hole and electron pockets at each calculated temperature. The overall shape of the distribution changes between 20 and 50K, and below 20K, the maximum values of the hole pocket lifetime distributions continue to increase, representing the emergence of high lifetime regions. c) Shows the ratio of the maximum and minimum values of $\tau_{e-ph}$ on the hole and electron pocket bands to further highlight the large variation of predicted lifetimes as well as the distinct difference in range between the hole and electron pockets. Importantly, we see a change between the calculated points at 20 and 50K, identifying the Fermi-surface linked features of hydrodynamic transport, which is reported to fully onset at 20K. In d) we visualize the appearance of the high lifetime regions across the Fermi surface with decreasing temperature. 
\label{fig:4}}
\end{figure*}

We instead predict that phonons dominate the momentum-conserving scattering processes. At temperatures below 100~K, overall electron-phonon scattering (with time $\tau\sub{e-ph}$)
is much faster than its momentum-relaxing component (with time $\tau\super{MR}\sub{e-ph}$).
This is because low energy acoustic phonons that dominate scattering at lower temperatures
predominantly change the electron momenta by small angles, as shown by the average
value of $\langle 1-\cos\theta\rangle$ dramatically reducing in magnitude with
decreasing temperature in Fig.~\ref{fig:2}(b).
Within a Debye model for the scattering time, $\tau\sub{e-ph} \propto T^{-4}$
and $\tau\super{MR}\sub{e-ph} \propto T^{-5}$ for $T \ll T_D$, the Debye temperature,
while both are $\propto T^{-1}$ for $T \gg T_D$.
Fig.~\ref{fig:2}(a) also shows the excellent Debye model fits to
the \emph{ab initio} predictions for $\tau\sub{e-ph}$ and $\tau\super{MR}\sub{e-ph}$,
confirming the acoustic phonon energy explanation for small-angle scattering.
Additionally, we find that phonon-mediated electron-electron scattering times also
substantially dominate over conventional Coulomb-mediated electron-electron scattering,
further confirming the central role played by phonons in the hydrodynamic behavior. 

We also evaluate the conductivity tensor from first principles, shown in Figure \ref{fig:3}, where we present the calculated resistivity overlaid with the corresponding experimental data, with mobility from Ref.~\citenum{gooth_electrical_2017} and resistivity from ~\citenum{schonemann_fermi_2017}. For both of these calculated transport properties (\emph{ab initio} resistivity and mobility), we find that our calculations are in excellent agreement with experimental reports presented as a function of temperature, indicating that our methods appropriately describes the microscopic properties of this material. 

To visualize the anisotropy of the electron-phonon lifetime across the Brillouin zone, we project the electron-phonon lifetime values across the Fermi surface.
The basic form of these surfaces is consistent with the `bowtie' electron  and quasi-2D `tube' hole pockets observed via ARPES.\cite{kumar_extremely_2017,razzoli_stable_2018} As shown in Fig. ~\ref{fig:4}a, there are two bands associated with each of the hole and electron pockets as a result of the strong spin-orbit coupling in WP\sub{2}. 

At low temperatures, the hole pockets, which determine the magnetotransport anisotropy for WP\sub{2},
feature particularly high lifetime regions in which $\tau\sub{e-ph}$ is on the scale of nanoseconds,
even microseconds at 2K, as shown in Fig ~\ref{fig:4}d.
High lifetimes over most of the Fermi surface at low temperatures 
is typical for conventional metals (see SI on copper and aluminum).
Particularly noteworthy for WP\sub{2} is the extreme anisotropy
spanning several orders of magnitude in the lifetimes,
with most of the Fermi surface remaining at $\sim 10^3-10^4$~fs
while certain spots exceed $10^8$~fs at 2~K.
we also include a series of distributions of $\tau\sub{e-ph}$ over the bands associated with the hole and electron pockets in Fig. ~\ref{fig:4}b,
as well as the ratio of the maximum and minimum lifetime values on the hole and electron pockets to quantify
the difference in the electron-phonon interaction strength across the Brillouin zone in in Fig. ~\ref{fig:4}c. 

These predictions present a unique microscopic perspective of hydrodynamic transport. The study of electron-phonon coupling and related scattering rates in WP\sub{2} shows that phonon related processes, rather than purely electron-electron processes, play a critical role in the emergence of hydrodynamic behavior. 

The calculated Fermi surface projected lifetimes show a distinct departure from behavior in traditional metals, making further investigation of electron-phonon effects near $E_F$ imperative. The prediction of severely anisotropic electron-phonon lifetimes at low temperatures is reminiscent of anisotropic electron-phonon models such as the ``cold spot'' model of cuprate superconductors. Understanding the full implications of this effect could strengthen the growing link between hydrodynamic materials and transport behavior in unconventional metals. This advance in descriptions of dynamics in WP\sub{2} opens a path to understanding transport effects in strongly correlated quantum systems and unconventional metals.

This work also serves as a foundation for further study of thermopower in hydrodynamic materials. Predictions of thermal conductivity and the contributions of the electronic and lattice components of thermopower could be used to understand the breakdown of the Wiedemann-Franz law. Going forward, the present methods could be extended to a fully first-principles framework for the identification of new hydrodynamic materials. 

\section*{Acknowledgements}
This research used resources of the National Energy Research Scientific Computing Center, a DOE Office of Science User Facility supported by the Office of Science
of the U.S. Department of Energy under Contract No. DE-AC02-05CH11231, as well as resources at the Research Computing Group at Harvard University.
JC recognizes the support of the DOE Computational Science Graduate Fellowship (CSGF).
RS acknowledges start-up funding from the Department of Materials Science and Engineering at Rensselaer Polytechnic Institute. PN acknowledges start-up funding from the Harvard John A. Paulson School of Engineering and Applied Sciences. The authors thank Amir Yacoby for fruitful discussions.

\section*{Author Information}
\subsection*{Contributions}
JC proposed the project and performed the calculations with PN. PN conceived the phonon mediated electron-electron scattering framework implemented by RS. PN supervised the project. All authors contributed to the analysis of the data and the writing of the manuscript.
\subsection*{Competing financial interests}
The authors declare no competing financial interests.

\makeatletter{} 
\end{document}